\documentstyle[12pt]{article}       
\begin{document}
%% collection of abbreviations %%%%%%%%%%%%%%%%%%%%%%%%%%%%%%%%%%%%%%%%%
\newcommand{\ncd}{\newcommand} \newcommand{\rcd}{\renewcommand}
%% LAYOUT %%%%%%%%%%%%%%%%%%%%%%%%%%%%%%%%%%%%%%%%%%%%%%%%%%%%%%%%%%%%%%
\ncd{\be}{\begin{equation}}        \ncd{\ee}{\end{equation}}
\ncd{\bq}{\begin{quote}}           \ncd{\eq}{\end{quote}}
%% TEXT %%%%%%%%%%%%%%%%%%%%%%%%%%%%%%%%%%%%%%%%%%%%%%%%%%%%%%%%%%%%%%%%
\ncd{\sps}{{super\-selec\-tion}}   \ncd{\st}{{sta\-tistics}}
\ncd{\dmn}{{dimension}}            \ncd{\obs}{{observable}}
\ncd{\thy}{{theory}}               \ncd{\spl}{{space-like}}
\ncd{\bgs}{{braid group \st}}      \ncd{\pgs}{{permutation group \st}}
\ncd{\rep}{{repre\-sen\-tation}}   \ncd{\vrep}{{vacuum \rep}}
\ncd{\irrep}{{irreducible \rep}}   \ncd{\per}{{positive-energy \rep}}
\ncd{\ca}{{current algebra}}       \ncd{\cf}{{con\-formal}}
\ncd{\trafo}{{trans\-for\-ma\-tion}}  \ncd{\qft}{{quantum field \thy}}
\ncd{\emt}{{energy-momentum tensor}}  \ncd{\vn}{von~Neumann}
\ncd{\comrel}{{commutation relation}} \ncd{\vna}{von~Neumann algebra}
%% MATH %%%%%%%%%%%%%%%%%%%%%%%%%%%%%%%%%%%%%%%%%%%%%%%%%%%%%%%%%%%%%%%%
\rcd{\aa}{{\cal A}}  \rcd{\d}{\delta}  \ncd{\po}{{\pi_0}}
\ncd{\NN}{{\sf I\!N}} \ncd{\ZZ}{{\sf Z\!\!Z}}
\ncd{\ainv}{\aa_{\rm inv}} \ncd{\avir}{\aa_{\rm Vir}}
%% HYPHEN %%%%%%%%%%%%%%%%%%%%%%%%%%%%%%%%%%%%%%%%%%%%%%%%%%%%%%%%%%%%%%
\hyphenation{mono-dromy}
%% BIBLIO %%%%%%%%%%%%%%%%%%%%%%%%%%%%%%%%%%%%%%%%%%%%%%%%%%%%%%%%%%%%%%
\def\CMP#1{Com\-mun.\ Math.\ Phys.\ {\bf #1}}
\def\RMP#1{Rev.\ Math.\ Phys.\ {\bf #1}}
\def\RH{R.\ Haag} \def\SD{S.\ Doplicher} \def\JR{J.E.\ Roberts}
\parindent0mm \parskip2mm
\rightline{DESY 93-115}
\rightline{August 1993}
{\Large \bf  News from the Virasoro Algebra}
\\[5mm]
{\large Karl-Henning Rehren}\\[3mm]
II.\ Inst.\ Theor.\ Physik, Univ.\ Hamburg, D-22761 Hamburg (FR Germany)
\\ {\small e-mail: i02reh@dsyibm.desy.de}
\addtolength{\baselineskip}{-1pt}
\bq
 {\bf Abstract:} It is shown that the local \qft\ of the chiral
 \emt\ with central charge $c=1$ coincides with the gauge invariant
 subtheory of the chiral $SU(2)$ \ca\ at level 1, where the gauge group
 is the global $SU(2)$ symmetry. At higher level, the same scheme gives
 rise to $W$-algebra extensions of the Virasoro algebra.
\eq
\addtolength{\baselineskip}{1pt} \parskip2mm

{\large \bf 1. Introduction}

Let $j^a(x)$ be the local currents of the chiral \ca\ of a compact
simple Lie group $G$ at level $k$, defined by the local \comrel s
\be -i[j^a(x),j^b(y)] = \sum_c f^{ab}_c \, j^c(x) \, \d(x-y) +
\frac k{2\pi} \, g^{ab} \, \d'(x-y). \ee
The metric $g^{ab}$ is normalized in terms of the structure constants
by $2h g^{ab} = - f^{ac}_d f^{bd}_c$ with $h$ the dual Coxeter number
of $G$ ($h = N$ for $G = SU(N)$). Let $T(x)$ be the \emt\ of the model
given by the Sugawara formula
\be T(x) = \frac \pi{k+h} \sum_{ab} g_{ab} : j^a(x)j^b(x) : \ee
and satisfying the \comrel s
\be -i[T(x),T(y)] = (T(x)+T(y)) \, \d'(x-y) - \frac c{24\pi} \,
\d'''(x-y) \ee
with central charge $c = dim(G) \cdot k/(k+h)$. The moments of $T(x)$
generate the \cf\ symmetry of the model, while the charges $Q^a = \int
j^a(x)dx$ generate a global $G$ symmetry. Under this symmetry, the
\emt\ is clearly invariant.

We emphasize the local form of the \comrel s rather than their Fourier
transforms, the Kac-Moody and Virasoro algebras. For $I \subset S^1$
an interval of the circle we denote by
\be \aa(I) \,\supset\, \ainv(I) \,\supset\, \avir(I) \ee
the local \vna s given by: the algebra $\{j^a(f) \,\vert\, {\rm supp}\,
f \subset I\}''$ associated with the currents (1) smeared with test
functions with support in the interval $I$, the subalgebra invariant
under $G$, and the algebra $\{T(f) \,\vert\, {\rm supp}\, f \subset
I\}''$ associated with the \emt\ (2),(3) smeared with test functions
with support in $I$, respectively. (The double primes denote the
von Neumann closure.) By locality, the algebras of disjoint
intervals commute \cite{BSM}. While the inclusions (4) are obvious, we
shall confirm a conjecture by B.Schroer and show that for $G = SU(2)$ at
level $k=1$
\be \avir(I) = \ainv(I). \ee
%\addtolength{\baselineskip}{.4pt} \parskip2mm

This result is in so far surprising as besides the \emt\ (2) also
bilocal expressions like
\be \sum_{ab} g_{ab} \, j^a(x)j^b(y)  \ee
are invariant. Indeed, it can be only true due to the weak closure of
the algebras (4), and does not hold for other groups or higher level.

The situation is reminiscent of an early similar result by Langerholc
and Schroer \cite{LS} who showed that the subalgebra of local
``currents'' $ : \varphi(x)\varphi(x) : $ in a four-dimensional theory
of free neutral bosons is not smaller than the even subalgebra
containing also bilocal operators $\varphi(x) \varphi(y)$. However, the
argument used by these authors crucially depends on the non-trivial
decay of the commutator function in light-like directions, and can
certainly not apply to the present situation with only $\d$-function
commutators (1). Indeed, as mentioned before, our finding is not
generic in chiral \qft\ but very specific to a particular model. \\[5mm]
{\large \bf 2. The main result}

The spectrum of the \cf\ Hamiltonian $L_0 = \frac 12 \int (1+x^2) T(x)
dx$ and the isospin component $Q^3$ is well known in the \per s $\pi$ of
the chiral $SU(2)$ \ca. It is described by the ``chiral partition
function''
\be \chi_\pi(q,t) := Tr_\pi(q^{Q^3}t^{L_0}) = \sum_{m,h} N_\pi(m,h)
q^m t^h \ee
where one may interprete $t = e^{-1/T}$ and $q = e^{-H}$ as a
``temperature'' and a ``magnetic field'' respectively. $N_\pi(m,h)$ are
the \dmn s of the eigenspaces $Q^3 \doteq m$, $L_0 \doteq h$ in the
\rep\ $\pi$. It is convenient to introduce the functions
\be [n]_q := \frac{q^{n/2}-q^{-n/2}}{q^{1/2}-q^{-1/2}} \ee
such that every contribution $[2j+1]_q = q^{-j} + q^{1-j} + \ldots +
q^{j-1} + q^j$ to the partition function corresponds to a full isospin
$j$ multiplet of states.

The irreducible \per s are determined by the \rep\ of $G$ in which the
subspace of lowest \cf\ energy transforms. For $SU(2)$ this is an
isospin $s = 0, \frac 12, \ldots \frac k2$. The corresponding chiral
partition functions are \cite{K,JF,KR}
\be \chi^{(k)}_s(q,t) = \frac
     {\sum_{n\in\ZZ} [2(k+2)n+2s+1]_q \cdot t^{h+n((k+2)n+2s+1)}}
     {\sum_{n\in\ZZ} [4n+1]_q \cdot t^{n(2n+1)}}  \ee
with $h = s(s+1)/(k+2)$. For the \vrep\ at level $k=1$ this is
\be \chi^{(1)}_0(q,t) = \frac
     {\sum_{n\in\ZZ} [6n+1]_q \cdot t^{n(3n+1)}}
     {\sum_{n\in\ZZ} [4n+1]_q \cdot t^{n(2n+1)}} . \ee
\bq
 {\bf Lemma 1:} The partition function (10) equals
 \be \chi^{(1)}_0(q,t) = \sum_{j\in\NN_0} [2j+1]_q \cdot
 t^{j^2}(1-t^{2j+1}) p(t). \ee
\eq
Here $p(t) = \prod_{n\in\NN} (1-t^n)^{-1}$. \pagebreak
Lemma 1 follows directly from \cite[Prop.\ 6.1]{S}.

Now, upon restriction to the $SU(2)$-invariant subalgebra $\ainv$, the
\vrep\ $\pi^0$ of the \ca\ $\aa$ decomposes into sub\rep s
$\pi^{\rm inv}_j$ characterized by the eigenvalue of the Casimir
operator $C \doteq j(j+1)$. Clearly, only integer isospins occur, and
every sub\rep\ $\pi^{\rm inv}_j$ arises with a full isospin $j$
multiplet as its multiplicity space. By inspection of Lemma 1, this
implies
\bq
 {\bf Lemma 2:} The partition function $\chi^{\rm inv}_j(t) =
 Tr_{\pi_j} t^{L_0}$ of the \rep\ $\pi^{\rm inv}_j$ of the gauge
 invariant subalgebra $\ainv$ (for $G = SU(2)$, $k=1$) is
 \be \chi^{\rm inv}_j(t) = t^{j^2}(1-t^{2j+1}) p(t). \ee
\eq
But (12) equals the partition function $\chi^{c=1}_h(t) = Tr_{\pi_h}
t^{L_0}$ of the \rep\ $\pi^{c=1}_h$ with lowest \cf\ energy $L_0 \doteq
h = j^2$ of the Virasoro algebra $\avir$ with central charge $c=1$,
given by the partition function $t^hp(t)$ of the Verma module
and a correction factor $(1-t^\nu)$ for the null space at level
$\nu = 2j+1$ \cite{KR}. Since $\avir \subset \ainv$, we conclude
\bq
 {\bf Lemma 3:} Upon restriction to the subalgebra $\avir$, the
 \irrep s $\pi^{\rm inv}_j$ of $\ainv$ remain irreducible and
 coincide with $\pi^{c=1}_h$, $h = j^2$.
\eq
In order to arrive at our main conclusion (5), we need one further
information. Namely, the local \qft\ of the \emt\ satisfies Haag
duality in the \vrep\ \cite{BSM}:
\bq
 {\bf Proposition 4:} In the \vrep\ $\pi_0$ of $\avir$
 \be \pi_0(\avir(I)) = \pi_0(\avir(I^c))'. \ee
\eq
Here $I^c$ denotes the complementary interval $S^1\setminus I$, and
the prime denotes the commutant of an algebra of operators. Now we are
ready to prove
\bq
 {\bf Proposition 5:} For $G = SU(2)$ and $k = 1$, one has
 \be \ainv(I) = \avir(I). \ee
\eq
{\it Proof:} Let $\pi_0$ be the \vrep\ of $\ainv$ which by Lemma 3
restricts to the \vrep\ of $\avir$. In the following chain of
inclusions, the first inclusion is evident, the second is due to
locality of $\ainv$, the third is the commutant of the first
(with $I^c$ instead of $I$), and the last equality is Prop.\ 4:
\be \pi_0(\avir(I)) \subset \pi_0(\ainv(I)) \subset \pi_0(\ainv(I^c))'
\subset \pi_0(\avir(I^c))' = \pi_0(\avir(I)). \ee
Since the \vrep\ is locally faithful, this implies Prop.\ 5.
\bq
 {\bf Corollary 6:} The \rep s $\pi^{c=1}_h$ of the \qft\ of the
 \emt\ with $c=1$ have statistical \dmn\ $d(\pi^{c=1}_h) = 2j+1$
 provided the lowest conformal energy is $h = j^2$, $j \in \NN_0$.
\eq
Namely, the statistical dimension of a sector of a gauge invariant
subtheory which is contained in the restriction of the \vrep\
(with unbroken symmetry) equals the dimension of the corresponding
\pagebreak
\rep\ of the symmetry group \cite{DHR}.

This seems to be the first non-trivial case that the statistical \dmn\
of a sector of the Virasoro algebra with $c \geq 1$ has been computed.
The value we found coincides with the asymptotic dimension defined by
the ``high-temperature'' limit $\frac 1T \searrow 0$, $t \nearrow 1$
\be d_{\rm as}(\pi) := \lim \frac{\chi_\pi(t)}{\chi_0(t)}. \ee
It is widely expected albeit never proven that $d(\pi) = d_{\rm
as}(\pi)$ holds quite generally for chiral theories. By \cite[Prop.\
8.2, 8.3]{KR}, this formula would yield infinite \st\ for all sectors
of $\avir$ with $c > 1$ except the vacuum sector (namely $\chi^c_h(t) =
t^h p(t)$ for $h > 0$ and $ = (1-t)p(t)$ for $h = 0$), as well as for
all sectors $\pi_h$ of $\avir$ with $c=1$ except when $h=j^2$ for $j \in
\frac 12 \NN_0$; in the latter case, $d(\pi_h) = 2j+1$. \\[5mm]
{\large \bf 3. Extended algebras}

One may repeat the analysis at higher level, or for other groups. One
has to expand the partition functions (9) (for $SU(2)$) of the \vrep\
into the form
\be \chi^{(k)}_0(q,t) = \sum_{j\in\NN_0} [2j+1]_q \cdot
\chi^{\rm inv}_j(t) \ee
and in turn expand $\chi^{\rm inv}_j(t)$ into the partition functions of
the corresponding Virasoro algebra with $c>1$. One finds that the
\irrep s of $\ainv$ no longer remain irreducible for $\avir$. E.g., at
level 2 one finds an infinite vacuum branching
\be \chi^{\rm inv}_0 = \chi^{c=3/2}_0 + \chi^{c=3/2}_4 + \chi^{c=3/2}_6
+ \chi^{c=3/2}_8 + \ldots \ee
showing that $\avir$ is a true subtheory of $\ainv$. Thus $\ainv$ is a
\cf ly invariant chiral theory containing apart from the \emt\ further
primary fields of scaling dimensions which can be read off the
character expansion. The level 2 example (18) exhibits the lowest lying
primary fields with dimensions 4, 6, and 8.

This is a general scheme to construct new local extensions
($W$-algebras) of the Virasoro algebra.

\rcd{\Large}{\large} \small \addtolength{\baselineskip}{-2pt}

\end{document}